\documentclass[preprint,aps]{revtex4-1}
\usepackage{graphics}
\usepackage{amsmath,amssymb}

\begin{document}

\baselineskip=20pt
\textheight=600pt

\title{Entanglement entropy bound in quantum mechanics}

\author{Maurizio Melis}
\email{E-mail address: maurizio.melis@gmail.com\\
Corresponding address: via Basilicata 31, 09127 Cagliari}

\bigskip

\begin{abstract}
We compute, in a quantum mechanical framework, the entanglement entropy of a spherically symmetric quantum system 
composed of two separate regions. In particular, we consider quantum states described by a wave function scale invariant and vanishing at infinity, with an asymptotic behaviour analogous to the case of a Coulomb potential or an harmonic oscillator. 

\noindent
We find that the entanglement entropy bound is proportional to the area of the boundary between the two 
separate regions, in accordance with the holographic bound on entropy. This study shows that 
the dependence of the maximum entanglement entropy on the boundary area can be derived in the context of quantum mechanics and does not necessarily require a quantum field theory approach.

\vspace{7 cm}


\bigskip


\end{abstract}

\maketitle

\section{Introduction}
\label{intro}

Einstein, Podolsky and Rosen (EPR) proposed a thought experiment 
\cite{EPR} 
to prove that quantum mechanics predicts the existence of ``spooky'' 
nonlocal correlations between 
spatially separate parts of a quantum system, a phenomenon that Schr\"odinger 
\cite{schrodinger} called {\it entanglement}. Afterward,
Bell \cite{bell} derived some inequalities that 
can be violated in quantum mechanics but must be satisfied 
by any local hidden variable model.
It was Aspect \cite{aspect} who first verified in laboratory that the EPR experiment, 
in the version proposed by Bohm \cite{bohm}, violates Bell inequalities, 
showing therefore 
that quantum entanglement and nonlocality are correct predictions of quantum mechanics.

A renewed interest in entanglement came from black hole physics: as suggested in 
\cite{sorkin,srednicki}, black hole entropy can be interpreted 
in terms of quantum entanglement, since the horizon of a black hole divides 
spacetime into two subsystems such that
observers outside cannot communicate the results of their 
measurements to observers inside, and vice versa. 
Black hole entanglement entropy turns out to scale with the area
${\cal A}$ of the event horizon, as supported by 
a simple argument proposed by Srednicki in \cite{srednicki}. 
If we trace over the field degrees of freedom located outside the black hole, 
the resulting entanglement entropy $S_A$ depend only on the degrees of freedom 
inside the black hole (regione $A$).  
If then we trace over the 
degrees of freedom inside the black hole, we obtain an entropy $S_B$ 
which depends only on the degrees of freedom outside (region $B$). 
It is straightforward to 
show that $S_A = S_B=S$, therefore the entropy $S$ should depend only 
on properties shared by the two regions inside and outside the black hole. The 
only feature they have in common is their boundary, so it is reasonable to expect 
that $S$ depends only on the area ${\cal A}$ of the event horizon. 

\noindent
This result is in accordance with the renowned Bekenstein-Hawking 
formula \cite{bekenstein1,bekenstein2,hawking1,hawking2}: 
$S_{BH}=\frac{\cal A}{4\ell_P^2}$, 
where $S_{BH}$ is the black hole entropy and $\ell_P$ is the Planck length. 

Some reviews and recent results on entanglement entropy in conformal field theory and  
black hole physics can be found in \cite{melis1,melis2,takayanagi,das}.
It is also a well-known property of many-body systems that the entanglement entropy obeys an ``area law'' 
with a logarithmic correction, as discussed e.g. in \cite{wolf,plenio,wolf2,cardy}. 

The area scaling of entanglement entropy has been investigated much more in the 
context of quantum field theory than in quantum mechanics. In order to bridge
the gap, we study the entanglement entropy of a quantum system composed 
of two separate parts, described by a 
wave function $\psi$, which we assume invariant under scale transformations and 
vanishing exponentially at infinity. We will show
that the entropy $S$ of both parts of our system is bounded by
$S \lesssim \eta \frac{\cal A}{4\ell_P^2}$, where $\ell_P$ is the Planck length and 
$\eta$ is a numerical constant related to the dimensionless parameter $\lambda$ 
appearing in the wave function $\psi$. 
This result, obtained at the leading order in $\lambda$, 
is in accordance with the so-called {\it holographic bound} 
on the entropy $S$ of an arbitrary system \cite{bousso,thooft,susskind}:
$S \leq \frac{\cal A}{4\ell_P^2}$, where 
$\cal A$ is the area of the surface enclosing the system. 
The holographic bound is an extension of the Bekenstein-Hawking formula for black hole entropy to all 
matter in the universe. 

In Section \ref{general} we present the main features of our approach, focusing in particular on the properties of 
entanglement entropy and on the form 
of the wave function describing the system. In Section \ref{results} we calculate analytically the bound on  
entanglement entropy. In Section \ref{conclusions} we summarize both the limits and the goals of our approach.  

\section{Main features of the model}
\label{general}
\begin{figure}[h]
\begin{center}
\resizebox{0.46\width}{0.46\height}{\includegraphics{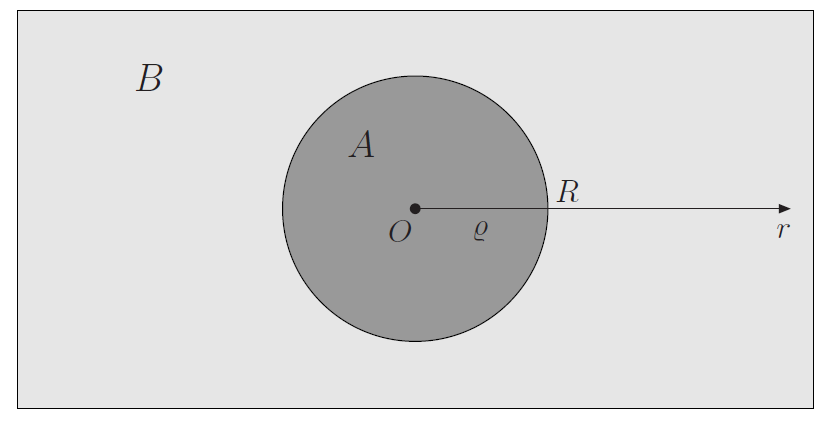}}
\caption{\label{regions}A quantum system composed of two parts 
$A$ and $B$, separated by a spherical surface of radius $R$.}
\end{center}
\end{figure}

Let us consider a spherically symmetric quantum system, 
composed of two regions $A$ and $B$ separated by a 
spherical surface of radius $R$ (Fig. \ref{regions}).

\noindent
The variables $\varrho$, $r$ describing the system  
are subjected to the following constraints (see Fig. \ref{regions}):
\begin{displaymath}
\left \{\begin{array}{l}
0\leq \varrho \leq R \quad\quad\, \mbox{region A}\\
r\geq R \quad\quad\quad\quad  \mbox{region B}\, ,
\end{array}
\right .
\end{displaymath}
where $R$ is the radius of the spherical surface separating the two regions. 
\noindent
It is convenient to introduce two dimensionless variables  
\begin{equation}
\label{coordinates}
x = \frac{\varrho}{R}, \quad \quad y = \frac{r}{R},
\end{equation}
subjected to the constraints $0\leq x \leq 1$ and $y\geq 1$. 

\noindent
In the following we will assume that the dynamics of the system is spherically symmetric, 
in order to treat the problem as one-dimensional in each region $A$ and $B$, with all 
physical properties depending only on the radial distance from the origin.

\subsection{Entanglement entropy}
\label{general1}
Let $\psi(x,\, y)$ be a generic wave function describing  
the system in Fig. \ref{regions}, composed of two parts $A$ and $B$. 

\noindent
As discussed e.g. in \cite{susskindbook,landau}, we can provide a  
description of all mesauraments 
in region $A$ through the so-called {\it density matrix} $\rho_{_A}(x,\, x')$:
\begin{equation}
\label{matrix1}
\rho_{_A}(x,\, x') = \int_{_B} d^3y\, \psi^*(x,\, y)\psi(x',\, y)\, ,
\end{equation}
where $d^3y$ is related to the volume element $d^3r$ in $B$ through the relation 
$d^3r = R^3d^3y$, with $d^3y = y^2\sin\theta d\theta\,d\phi\,dy$ 
in spherical coordinates.  

\noindent
Similarly, experiments performed in $B$ are described by 
the density matrix $\rho_{_B}(y,\, y')$:
\begin{equation}
\label{matrix2}
\rho_{_B}(y,\, y') = \int_{_A} d^3x\, \psi^*(x,\, y)\psi(x,\, y')\, ,
\end{equation}
where $d^3x$ is related to the volume element $d^3\varrho$ in $A$ through the relation 
$d^3\varrho = R^3d^3x$, with $d^3x = x^2\sin\theta d\theta\,d\phi\,dx$ 
in spherical coordinates.

\noindent
Notice that $\rho_{_A}$ is calculated tracing out the exterior variables $y$, whereas 
$\rho_{_B}$ is evaluated tracing out the interior variables $x$.

\noindent
Density matrices have the following properties:
\begin{enumerate}
\item Tr$\, \rho = 1$ (total probability equal to 1)
\item $\rho = \rho^{\dagger}$ (hermiticity)
\item $\rho_j \geq 0$ (all eigenvalues are positive or zero).
\end{enumerate} 

\noindent
When only one eigenvalue of $\rho$ is nonzero, the nonvanishing 
eigenvalue is equal to 1 by virtue of the trace condition on $\rho$.
This case occurs only if the wave function can be factorized into an 
uncorrelated product 
\begin{equation}
\label{uncorrelated}
\psi(x,\, y) = \psi_{_A}(x)\cdot\psi_{_B}(y)\, .
\end{equation}

\noindent
A quantitative measure of the degree of entanglement between 
the two parts $A$ and $B$ of the system is provided by the {\it von Neumann entropy}
\begin{equation}
\label{ee}
S = -\mbox{Tr}(\rho\ln\rho)\, ,
\end{equation}
which is also called {\it entanglement entropy}.
 
\noindent
When the two subsystems $A$ and $B$ are each the complement of the other, 
entanglement entropy can be calculated by tracing out the variables associated to 
region $A$ 
or equivalently to region $B$, since it turns out $S_{_A}=S_{_B}$. 

\subsection{Wave function}
\label{general2}
The spherical region $A$ in Fig. \ref{regions} is part of a larger 
closed system $A\cup B$, described by a wave function $\psi (x,\, y)$, 
where $x$ denotes the set of coordinates in $A$ and 
$y$ the coordinates in $B$. 
We will exploit for $\psi$ the following analytic forms:
\begin{equation}
\label{wave}
\psi (x,\, y) = C_n\, e^{-\lambda y^n/x^n} \quad\quad \mbox{(with}\;\; n=1 \;\;  \mbox{or}\;\; n=2\,\mbox{)}\, ,  
\end{equation}
where $C_n$ is the normalization constant and $\lambda$ is a dimensionless parameter, 
which we assume much greater than unity ($\lambda \gg 1$).

\noindent
From the point of view of an observer in $A$, the asymptotic behaviour of $\psi (x,\, y)$, for $n=1$, is an exponential decay, as in the case of a Coulomb potential, while for $n=2$ the wave function $\psi$ has an asymptotic gaussian slope, as in the case of an harmonic oscillator.

\noindent
If the system is in a quantum state subjected to a central potential,  
the complete wave function $\Phi$ 
should contain an angular part expressed by spherical harmonics 
$Y_{lm}(\theta,\, \phi)$:
\begin{equation}
\Phi (x,\,y;\,\theta,\,\phi) = Y_{lm}(\theta,\, \phi)\,\psi (x,\, y)\, , 
\end{equation}
If the angular momentum is zero, the angular component of the wave function reduces to a constant 
$Y_{00}(\theta,\, \phi) = 1/\sqrt{4\pi}$ and can be included in the constant $C_n$ [Eq. (\ref{wave})], which 
hence represents the overall normalization constant of the complete wave function $\Phi$. 

\noindent
In order to justify the form (\ref{wave}) of the wave function $\psi$, let us list 
the main properties it satisfies:

\smallskip

\noindent
{\bf 1)} $\psi$ depends on both sets of variables $x,\, y$ 
defined in the two separate regions 
$A$ and $B$, but it is not factorizable in the product (\ref{uncorrelated}) 
of two distinct parts 
depending only on one variable: 
\begin{equation}
\psi (x,\, y) \neq \psi_{_A}(x)\cdot \psi_{_B}(y)\, .
\end{equation}
This assumption guarantees that the entanglement entropy of the system 
is not identically zero.

\smallskip

\noindent
{\bf 2)} $\psi$ depends on the variables $x,\, y$ through their ratio $y/x$, hence the wave function 
is invariant under scale transformations: 
\begin{equation}
x\to\mu x \;\; \mbox{and} \;\; y\to\mu y, \;\; \mbox{with} \;\; \mu \;\; 
\mbox{constant}\, .  
\end{equation}

\smallskip

\noindent
{\bf 3)} $\psi$ has the asymptotic form of the wave function for a quantum state in a Coulomb potential (for $n=1$) 
or for an harmonic oscillator (for $n=2$).

\noindent
The asymptotic form of the radial part $f(r)$ of the wave functions describing the quantum states in 
a Coulomb potential or for an harmonic oscillator is
\begin{equation}
\label{erre}
f(r) \approx e^{-\kappa r^n} \approx e^{-\kappa R^n y^n}\, , \quad \mbox{with}\; n=1\; \mbox{or}\; n=2\, .
\end{equation}
$r\to\infty$ is the radial distance from the origin, $y$ is defined by $y=r/R$ while the parameter $\kappa$ is 
\begin{equation}
\label{kappa}
\kappa = \frac{\sqrt{2m|E|}}{n\hbar R^{n-1}}
\end{equation} 
for a particle with mass $m$ and energy $E$ ($\hbar$ is the Planck constant divided by $2\pi$). 

\noindent
The Coulomb case ($n=1$) corresponds to a bound state with negative energy ($E<0$), while the harmonic oscillator
case ($n=2$) has positive energy ($E>0$).

\noindent
The radial part $f(r)$ of the wave function coincides with 
the restriction $\psi_{_B}(y)$ of the wave function $\psi(x,\,y)$ to the 
exterior region $B$, as seen by 
an inner observer localized for instance at $x=1$, i.e. on  
the boundary between the two regions:
\begin{equation}
\label{restriction}
\left. \psi_{_B}(y)=\psi(x,\,y)\right|_{x=1} \approx e^{-\lambda y^n}\, .
\end{equation}
By comparing the asymptotic behaviour of the wave functions 
(\ref{erre}) and (\ref{restriction}) as $y\to\infty$, we find:
\begin{equation}
\label{lambdalast}
\lambda = \gamma\,R\, , \quad \mbox{with} \quad 
\gamma = \frac{\sqrt{2m|E|}}{n\hbar}\, .
\end{equation}
In Section \ref{results} we will assume $\lambda\gg 1$, which is always true in a 
system with $R$ sufficiently large, as it easily follows from 
Eq. (\ref{lambdalast}). 


\noindent
If the inner observer is not localized on the boundary 
but in a fixed point $x_0$ inside region $A$ (with $0<x_0<1$), then the 
expression (\ref{lambdalast}) of $\lambda$ has to be multiplied by $x_0$. 

\noindent
Notice that the dependence of $\lambda$ on the radius $R$ of the boundary 
has been derived by imposing an asymptotic form  
on the wave function $\psi(x,\,y)$ as $y\to\infty$, with respect to 
a fixed point $x\sim 1$ inside the spherical region of the system in Fig. \ref{regions}.

In Section \ref{results} we will show that the entropy of both parts of our system 
depends on  $\lambda^2$, which in turn is proportional to the area ${\cal A}$ of the spherical boundary. 
The area scaling of the entanglement entropy 
is, essentially, a consequence of the nonlocality of the wavefunction $\psi(x,\,y)$, 
which establishes a correlation 
between points inside ($x\sim 1$) and outside ($y\to\infty$) the boundary.

\section{Analytic results}
\label{results}

We normalize the wave function (\ref{wave}) by imposing the condition  
\begin{equation}
\int_{_A} d^3x\int_{_B} d^3y\, \psi^*(x,\, y)\psi (x,\, y) = 1\, ,
\end{equation}
with  $d^3x = x^2\sin\theta d\theta\,d\phi\,dx$ and 
$d^3y = y^2\sin\theta d\theta\,d\phi\,dy$ in spherical coordinates. 
Under the assumption 
$\lambda\gg 1$, the normalization constant $C_n$ in Eq. (\ref{wave}) turns out to be
\begin{equation}
\label{normalization}
C_n \approx 2^{n-1}\lambda\frac{e^{\lambda}}{\sqrt{\pi}}\, ,
\end{equation}
as easily obtained by means of Wolfram Mathematica 11.1 \cite{math}.

\noindent
Let us focus on the interior region $A$ represented in Fig. \ref{regions}. 
We calculate the density matrix $\rho_{_A}(x,\, x')$ by tracing out the variables $y$ 
related to the subsystem $B$, as expressed in Eq. (\ref{matrix1}):
\begin{eqnarray}
\label{density}
\rho_{_A}(x,\, x') & = & \int_{_B} d^3y\, \psi^*(x,\, y)\psi (x',\, y) \nonumber \\
& \approx &  \frac{C_n^2}{2^{n-1}\lambda}\frac{x^nx'^n}{x^n+x'^n}e^{-\lambda\frac{x^n+x'^n}{(xx')^n}}\, ,  
\end{eqnarray}
where we assumed $\lambda\gg 1$.

\noindent
It is easy to verify that the density matrix $\rho_{_A}(x,\, x')$ satisfies all 
properties listed in \ref{general1}:
\begin{enumerate}
\item Total probability equal to 1:
\begin{displaymath}
\int_{_A} d^3x\, \rho_{_A}(x,\, x) = 1 \Longleftrightarrow \mbox{Tr}(\rho_{_A}) = 1\, . 
\end{displaymath}
\item Hermiticity:
\begin{displaymath}
\rho_{_A}(x,\, x') = \rho^*_{_A}(x',\, x) \Longleftrightarrow \rho_{_A} = 
\rho_{_A}^{\dagger}\, .
\end{displaymath}
\item All eigenvalues are positive or zero:
\begin{displaymath}
\rho_{_A}(x,\, x') \geq 0 \quad \forall\; x,\, x'\in (0,\, 1) \Longrightarrow 
\big(\rho_{_A}\big)_j \geq 0 \, .
\end{displaymath}
\end{enumerate}
The entanglement entropy (\ref{ee}) can be expressed in the form
\begin{equation}
\label{entropy}
S_A = -\int_{_A} d^3x\, \rho_{_A}(x,\, x)\ln[\rho_{_A}(x,\, x)]\, . 
\end{equation}
Substituting the expression (\ref{density}) of $\rho_{_A}(x,\, x')$, 
with $x'=x$, we find: 
\begin{equation}
S_A \approx \frac{C_n^2}{2^n\lambda}
\int_{_A} d^3x\, e^{-2\lambda/x^n}x^n\ln\left(\frac{C_n^2}{2^n\lambda}\,e^{-2\lambda/x^n}x^n\right )\, . 
\end{equation}
We can maximize the previous integral by means of the condition
\begin{equation}
e^{-2\lambda/x} \leq e^{-2\lambda} \quad \forall\; x \in (0,\, 1)\, .
\end{equation}
By neglecting the subleading terms in $\lambda\gg 1$ and inserting the expression (\ref{normalization}) of the normalization constant $C_n$, 
the entanglement entropy $S_A$ turns out to be bounded by
\begin{equation}
S_A \lesssim \left(\frac{16}{5}\right)^{n-1}\frac{\lambda^2}{3}\, .
\end{equation}
If we substitute $\lambda=\gamma\,R$, as given in Eq. (\ref{lambdalast}), we finally find:
\begin{equation}
\label{final}
S_A \lesssim \eta\,\frac{\cal A}{4\ell^2_P}\, , \quad \mbox{with} \quad 
\eta = \left(\frac{16}{5}\right)^{n-1}\frac{\ell^2_P}{3\pi}\,\gamma^2\, ,
\end{equation}
where ${\cal A}=4\pi R^2$ is the area of the spherical boundary in Fig. \ref{regions}, 
$\ell_P = \left(\hbar G/c^3\right)^{1/2}$ is the Planck length and $\gamma$ is given in Eq. (\ref{lambdalast}).
The result (\ref{final}) is in accordance with the holographic bound on entropy
$S\leq\frac{\cal A}{4\ell^2_P}$,  
discussed in \cite{bousso,thooft,susskind}, and 
shows that the entanglement entropy of both parts of our 
composite system obeys an ``area law''.

\noindent
For a particle with energy $E$ and mass $m$ we can express the parameter $\eta$ in the form 
\begin{equation}
\label{eta}
\eta = \left(\frac{16}{5}\right)^{n-1}\frac{2}{3\pi}\,\frac{m|E|}{n\, m_P^2 c^2}\, , 
\end{equation}
where we combined Eqs. (\ref{lambdalast}), (\ref{final}) and introduced the Planck mass 
$m_{_{P}} = (\hbar c/G)^{1/2}$. 
Under the assumptions $|E|\ll m_Pc^2$ and $m\lesssim m_{_{P}}$, we obtain
$\eta\ll 1$, therefore in this case the bound (\ref{final}) on entropy is 
much stronger than the holographic bound $S\leq\frac{\cal A}{4\ell^2_P}$.
For instance, if we consider the electron in the ground state of the hydrogen atom, we find 
$\eta = \frac{2}{3\pi}\,\frac{mc^2|E|}{(m_P c^2)^2} \approx 10^{-50}$, being $mc^2 = 0.511$ MeV, 
$|E| = 13.6$ eV and $m_{_{P}}c^2 \approx 1.2\cdot 10^{19}$ GeV.

All calculations performed in this Section can be repeated by focusing on the 
exterior region $B$ represented in Fig. \ref{regions}. If we trace out the interior 
variables $x$, as in Eq. (\ref{matrix2}), the density matrix 
$\rho_{_B}(y,\, y')$ turns out to be 
\begin{eqnarray}
\rho_{_B}(y,\, y') & = & \int_{_A} d^3x\, \psi^*(x,\, y)\psi (x,\, y')\nonumber \\ 
& \approx & \frac{C_n^2}{2^{n-1}\lambda}\frac{e^{-\lambda(y^n+y'^n)}}{y^n+y'^n}\quad \mbox{(with}\; n=1\; 
\mbox{or}\; n=2\mbox{)}\, , 
\end{eqnarray}
where we substituted the expression (\ref{wave}) of the wave function $\psi$  
and applied the usual assumption $\lambda\gg 1$.

\section{Conclusions}
\label{conclusions}
In this study we proposed a simple approach to the calculation of the entanglement entropy of
a spherically symmetric quantum system. 
The result obtained in Eq. (\ref{final}), $S_A \lesssim \eta\,\frac{\cal A}{4\ell^2_P}$, 
is in accordance with the holographic 
bound on entropy and with the ``area law'' discussed e.g. in \cite{wolf,plenio,wolf2}.
Our result, derived in the context of quantum mechanics, shows that the maximum entanglement entropy of a system  is proportional to the area of the boundary and not to the volume of the system, as one would have expected. 
Therefore, it turns out that the information content 
of a system is stored on its boundary rather than in the bulk \cite{susskindbook}.

\noindent
The area scaling of the entanglement entropy 
is a consequence of the nonlocality of the 
wave function, which relates the points inside the boundary 
with those outside. In particular, we derived the area law for entropy by 
imposing an asymptotic behaviour on the wave function $\psi(x,\,y)$ as $y\to\infty$, 
with respect to a fixed point $x\sim 1$ inside the interior region.

\noindent
The main limit of our model is that we considered only two particular forms of  
the wave function $\psi$. However, more general forms of $\psi$ 
might be considered in future developments of the model. 
Let us finally stress
that our results are valid 
at the leading order in the dimensionless parameter $\lambda\gg 1$ appearing in 
the wave function $\psi$ of the system.

\noindent
The treatment presented in this study 
for the entanglement entropy of a quantum system is an extremely 
simplified model, 
but the accordance of our result with the holographic bound 
and with the area scaling of the entanglement entropy indicates that we  
isolated the essential physics of the problem in spite of all simplifications.

\noindent
A remarkable finding of this study is also that the area scaling of the entanglement entropy bound is intrinsic to quantum mechanics and does not necessarily depend on quantum field theory.

\begin{acknowledgments}
\noindent
Most of the work in this study was carried out during the author's Ph.D. in Nuclear and Subnuclear Physics and Astrophysics at Physics Department of Cagliari University. 

\noindent
The author is very greatful to his supervisor, Prof. M. Cadoni, for helpful discussions and criticism.
\end{acknowledgments}

\appendix*

\section{Nonlocal potentials}
\noindent
In order to calculate the nonlocal potential $V(x,\,y)$, 
corresponding to the wave function $\psi (x,\, y)$ in Eq. (\ref{wave}),
let us consider the time-independent Schr\"odinger equation
\begin{displaymath}
\left(-\frac{\hbar^2}{2m}\nabla^2+V\right)\psi(x,\,y) = E\psi(x,\,y)\, ,
\end{displaymath}
where $E$ is the ground state energy of the system while
the Laplace operator $\nabla^2$ is expressed by the identity
\begin{displaymath}
\nabla^2 \equiv \frac{1}{r^2}\frac{\partial}{\partial r}\left 
(r^2\frac{\partial}{\partial r}\right ) \equiv 
\frac{2}{r}\frac{\partial}{\partial r} + \frac{\partial^2}{\partial r^2}\, ,
\end{displaymath}
valid for spherical symmetric systems with radial coordinate $r$.

\noindent
The restrictions of the potential $V(x,\,y)$ to the regions $A$ and $B$ are defined, 
respectively, as
\begin{displaymath}
V_{_{A}}(x) = \left . V_{_{A}}(x,\, y)\right |_{y=1} \quad \mbox{and} \quad 
V_{_{B}}(y) = \left . V_{_{B}}(x,\, y)\right |_{x=1}\, .
\end{displaymath}

\noindent
Analogously, the wave function $\psi (x,\, y)$ in the regions $A$ and $B$ of our system can be plotted by fixing 
one of the two variables $x$, $y$ and representing $\psi$ with respect 
to the other variable.

\subsection{Asymptotic Coulomb potential}
By setting $y=1$ or $x=1$ in Eq. (\ref{wave}) with $n=1$ (in the case of an asymptotic 
Coulomb potential), we obtain 
the following restrictions of $\psi (x,\, y)$ to the regions $A$ or $B$ of our system (Fig. 1):
\begin{eqnarray}
\psi_{_A}(x) & = & \left .\psi(x,\, y)\right |_{y=1} = Ce^{-\frac{\lambda}{x}}\, , 
\quad \mbox{with} \; x\in (0,\, 1) \nonumber \\
\psi_{_B}(y) & = & \left .\psi(x,\, y)\right |_{x=1} = Ce^{-\lambda y}\, , \quad  
\mbox{with} \; y\in (1,\, +\infty)\, .\nonumber
\end{eqnarray}
The shape of $\psi_{_{A}}(x)$ and $\psi_{_{B}}(y)$ is represented in the following figure:
\begin{displaymath}
\resizebox{0.52\width}{0.52\height}{\includegraphics{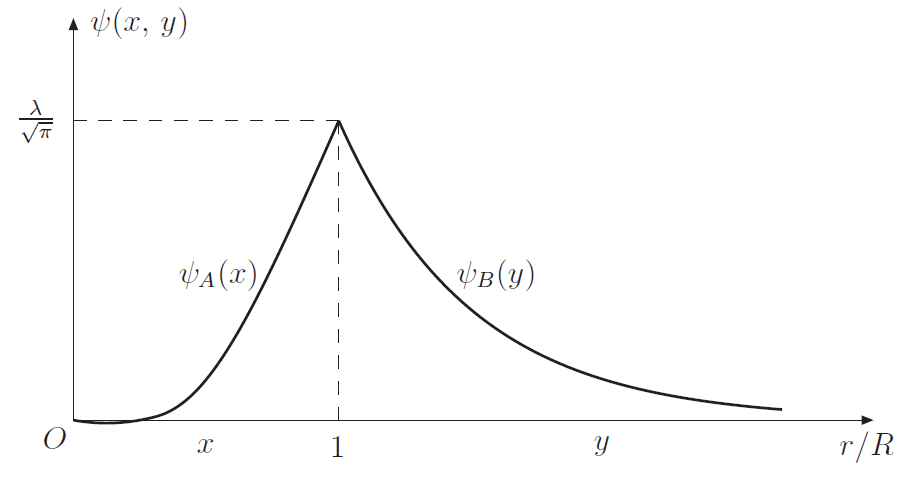}}
\end{displaymath}

\noindent
In region $A$ we obtain
\begin{displaymath}
V_{_{A}}(x,\, y) = E_{_{A}} + \frac{\hbar^2\lambda^2}{2mR^2}\frac{y^2}{x^4}\, ,
\end{displaymath}
while in region $B$ we find
\begin{displaymath}
V_{_{B}}(x,\, y) = E_{_{B}} + \frac{\hbar^2\lambda^2}{2mR^2}\frac{1}{x^2}
\left(1-\frac{2x}{\lambda y}\right)\, .
\end{displaymath}
By setting the zero of the potential $V(x,\,y)$ at infinity (for $y\to\infty$), 
we obtain
\begin{displaymath}
\lim_{y\to\infty}V_{_{B}}(y) = 0 \Longrightarrow 
E_{_{B}} = -\frac{\hbar^2\lambda^2}{2mR^2}\, .
\end{displaymath}
By imposing the continuity of the potential $V(x,\,y)$ on the boundary 
(for $x=y=1$), we find
\begin{displaymath}
\left . V_{_{A}}(x)\right |_{x=1} = \left . V_{_{B}}(y)\right |_{y=1} 
\Longrightarrow E_{_{A}} = -\frac{\hbar^2\lambda^2}{2mR^2}
\left(1+\frac{2}{\lambda}\right)\, .
\end{displaymath}
The restriction of the potential $V(x,\,y)$ to the region $A$ is
\begin{displaymath}
V_{_{A}}(x) = \frac{\hbar^2\lambda^2}{2mR^2}\left(\frac{1}{x^4}-1-
\frac{2}{\lambda}\right)\quad \mbox{with} \; x\in (0,\, 1)\, ,
\end{displaymath}
while the restriction of $V(x,\,y)$ to the region $B$ is
\begin{displaymath}
V_{_{B}}(y) = -\frac{\hbar^2\lambda}{mR^2}\frac{1}{y}
\quad \mbox{with} \; y\in (1,\, +\infty)\, .
\end{displaymath}
The behaviour of $V_{_{A}}(x)$ and $V_{_{B}}(y)$ is represented in the following figure:
\begin{displaymath}
\resizebox{0.42\width}{0.42\height}{\includegraphics{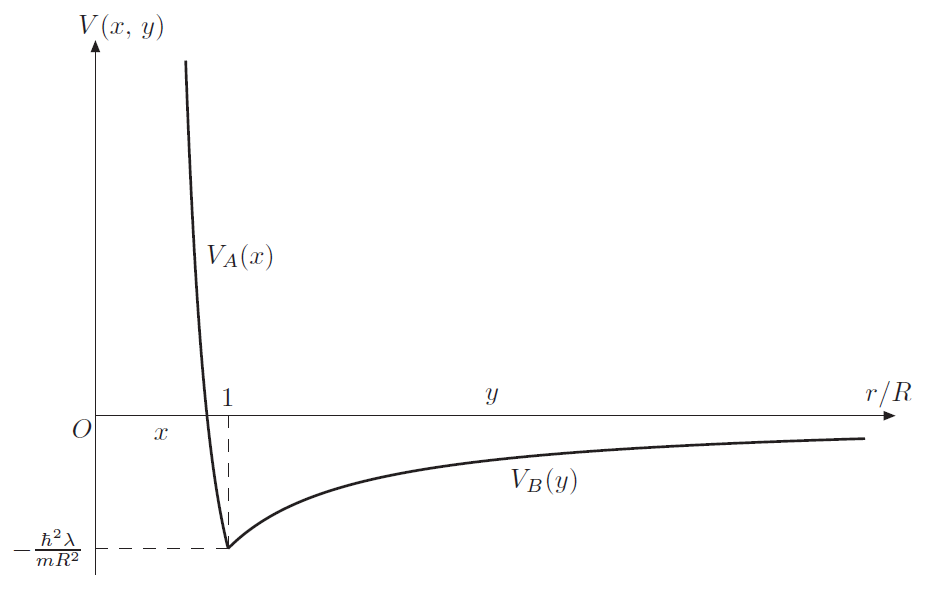}}
\end{displaymath}
Finally, the nonlocal potential $V(x,\,y)$, 
corresponding to the nonlocal wave function $\psi (x,\, y)$ [Eq. (\ref{wave}) with $n=1$] is given by
\begin{displaymath}
V(x,\,y) = \left\{\begin{array}{ll}
\frac{\hbar^2\lambda^2}{2mR^2}\left(\frac{y^2}{x^4}-1-\frac{2}{\lambda}\right) 
\quad & \mbox{region}\; A\\
\frac{\hbar^2\lambda^2}{2mR^2}\left(\frac{1}{x^2}-\frac{2}{\lambda xy}-1\right)
\quad & \mbox{region}\; B\, .
\end{array}\right . 
\end{displaymath}

\subsection{Asymptotic harmonic oscillator potential}
By setting $y=1$ or $x=1$ in Eq. (\ref{wave}) with $n=2$ (in the case of an asymptotic harmonic oscillator potential), we obtain 
the following restrictions of $\psi (x,\, y)$ to the regions $A$ or $B$ of our system (Fig. 1):
\begin{eqnarray}
\psi_{_A}(x) & = & \left .\psi(x,\, y)\right |_{y=1} = Ce^{-\frac{\lambda}{x^2}}\, , 
\quad \mbox{with} \; x\in (0,\, 1) \nonumber \\
\psi_{_B}(y) & = & \left .\psi(x,\, y)\right |_{x=1} = Ce^{-\lambda y^2}\, , \quad  
\mbox{with} \; y\in (1,\, +\infty)\, .\nonumber
\end{eqnarray}
The shape of $\psi_{_{A}}(x)$ and $\psi_{_{B}}(y)$ is represented in the following figure:
\begin{displaymath}
\resizebox{0.42\width}{0.42\height}{\includegraphics{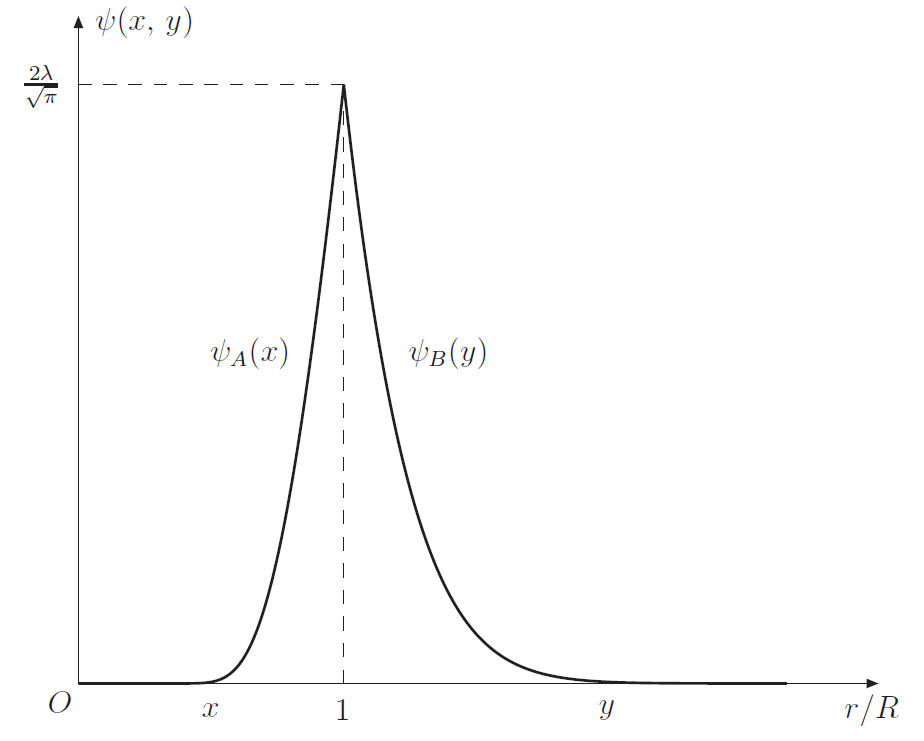}}
\end{displaymath}

\noindent
In region $A$ we obtain
\begin{displaymath}
V_{_{A}}(x,\, y) = E_{_{A}} + \frac{\hbar^2\lambda}{mR^2}\frac{1}{x^4}\left(\frac{2\lambda}{x^2}-1\right)\, ,
\end{displaymath}
while in region $B$ we find
\begin{displaymath}
V_{_{B}}(x,\, y) = E_{_{B}} + \frac{\hbar^2\lambda}{mR^2}\left(2\lambda y^2-3\right)\, .
\end{displaymath}
By setting the zero of the potential $V(x,\,y)$ on the boundary (for $x=y=1$),  
we obtain
\begin{displaymath}
\left . V_{_{A}}(x)\right |_{x=1} = 0 \Longrightarrow E_{_{A}} = \frac{\hbar^2\lambda}{mR^2}\left(1-2\lambda\right)
\end{displaymath}
\begin{displaymath}
\left . V_{_{B}}(y)\right |_{y=1} = 0 \Longrightarrow
E_{_{B}} = \frac{\hbar^2\lambda}{mR^2}\left(3-2\lambda\right)\, .
\end{displaymath}
The restriction of the potential $V(x,\,y)$ to the region $A$ is
\begin{displaymath}
V_{_{A}}(x) = \frac{\hbar^2\lambda}{mR^2}\left[\frac{1}{x^4}\left(\frac{2\lambda}{x^2}-1\right)+1-2\lambda\right]
\quad \mbox{with} \; x\in (0,\, 1)\, ,
\end{displaymath}
while the restriction of $V(x,\,y)$ to the region $B$ is
\begin{displaymath}
V_{_{B}}(y) = \frac{2\hbar^2\lambda^2}{mR^2}\left(y^2-1\right)\quad \mbox{with} \; y\in (1,\, +\infty)\, .
\end{displaymath}
The behaviour of $V_{_{A}}(x)$ and $V_{_{B}}(y)$ is represented in the following figure:
\begin{displaymath}
\resizebox{0.42\width}{0.42\height}{\includegraphics{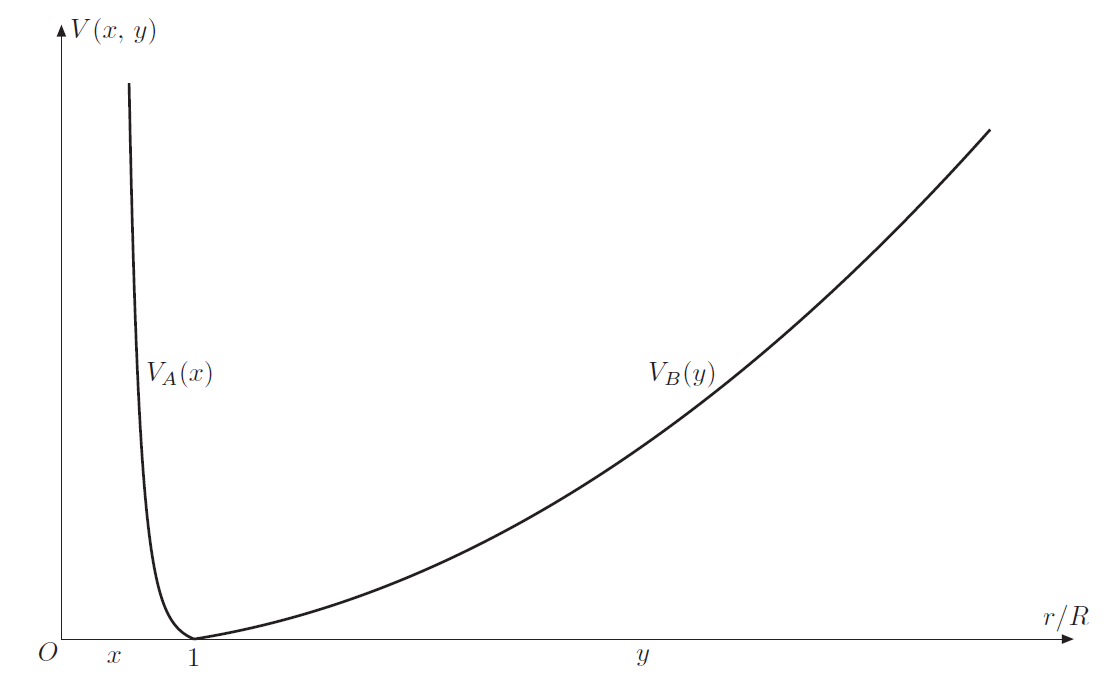}}
\end{displaymath}

\noindent
Finally, the nonlocal potential $V(x,\,y)$, 
corresponding to the nonlocal wave function $\psi (x,\, y)$ [Eq. (\ref{wave}) with $n=2$] is given by
\begin{displaymath}
V(x,\,y) = \left\{\begin{array}{ll}
\frac{\hbar^2\lambda}{mR^2}\left[\frac{y^2}{x^4}\left(\frac{2\lambda y^2}{x^2}-1\right)+1-2\lambda\right] 
\quad & \mbox{region}\; A\\
\frac{\hbar^2\lambda}{mR^2}\left[\frac{1}{x^2}\left(\frac{2\lambda y^2}{x^2}-3\right)+3-2\lambda\right]
\quad & \mbox{region}\; B\, .
\end{array}\right . 
\end{displaymath}

\end{document}